\documentclass{article}
\usepackage{arxiv}
\usepackage{booktabs} 
\usepackage{amsmath}
\usepackage{algorithm}
\usepackage{listings}
\usepackage{algorithmic}
\usepackage{xcolor}
\usepackage{multirow}
\usepackage{xcolor}
\definecolor{cardinal} {RGB}{196, 30, 58}
\definecolor{lightgrey}{RGB}{150,150,150}
\usepackage{enumitem}
\usepackage{bbding}
\usepackage{diagbox}
\usepackage{graphicx}
\usepackage{pifont}
\usepackage{hyperref}
\usepackage[skipempty]{credits}

\usepackage{xcolor}
\definecolor{mygreen}{HTML}{548235}
\definecolor{myorange}{HTML}{C55A11}

\usepackage{balance}

\title{LogBabylon: A Unified Framework for Cross-Log File Integration and Analysis}

\author{\href{https://orcid.org/0000-0000-0000-0000}{\includegraphics[scale=0.06]{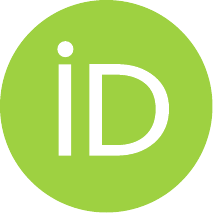}\hspace{1mm}Rabimba Karanjai}\footnotemark[1] \\
	Department of Computer Science\\
	University of Houston\\
	Houston, TX, United States \\
	\texttt{rabimba@cs.uh.edu} \\
    \And
    Yang Lu\thanks{Both authors contributed equally to this research and are equal first authors.} \\
	Department of Computer Science\\
	University of Houston\\
	Houston, TX, United States \\
	\texttt{ylu17@central.uh.edu} \\
	\And
	Dana Alsagheer \\
	Department of Computer Science\\
	University of Houston\\
	Houston, TX, United States \\
	\And
	Keshav Kasichainula \\
	Department of Computer Science\\
	University of Houston\\
	Houston, TX, United States \\
	\And
	Lei Xu \\
	Department of Computer Science\\
	Kent State University\\
	Kent, OH, United States \\
	\And
	Weidong Shi \\
	Department of Computer Science\\
	University of Houston\\
	Houston, TX, United States \\
	\And
	Shou-Hsuan Stephen Huang \\
	Department of Computer Science\\
	University of Houston\\
	Houston, TX, United States \\
}

\credit{Rabimba}{Conceptualization, Methodology,Investigation, Software,Data curation,Formal analysis,  Writing -- original draft}
\credit{Yang}{Conceptualization, Methodology, Investigation, Writing -- review \& editing, Visualization}
\credit{Dana}{Data curation, Writing -- review \& editing}
\credit{Keshav}{Data curation, Writing -- review \& editing}
\credit{Lei}{Data curation, Investigation, Software, Writing -- original draft}
\credit{Weidong}{Data curation, Writing -- original draft, Visualization, Funding acquisition}
\credit{Stephan}{Funding acquisition}

\begin{document}

\maketitle
\begin{abstract}
Logs are critical resources that record events, activities, or messages produced by software applications, operating systems, servers, and network devices. However, consolidating the heterogeneous logs and cross-referencing them is challenging and complicated. Manually analyzing the log data is time-consuming and prone to errors. LogBabylon is a centralized log data consolidating solution that leverages Large Language Models (LLMs) integrated with Retrieval-Augmented Generation (RAG) technology. LogBabylon interprets the log data in a human-readable way and adds insight analysis of the system performance and anomaly alerts. It provides a paramount view of the system landscape, enabling proactive management and rapid incident response. LogBabylon consolidates diverse log sources and enhances the extracted information's accuracy and relevancy. This facilitates a deeper understanding of log data, supporting more effective decision-making and operational efficiency. Furthermore, LogBabylon streamlines the log analysis process, significantly reducing the time and effort required to interpret complex datasets. Its capabilities extend to generating context-aware insights, offering an invaluable tool for continuous monitoring, performance optimization, and security assurance in dynamic computing environments.
\end{abstract}
\section{Introduction}
In today's large-scale production systems, log collection plays a crucial role as a source of valuable information. Logs provide key insights into a system's operational status, enabling problem identification, resolution, and performance optimization. However, logs vary in format and content depending on the system or application that generates them. Typically, they consist of unstructured text produced by logging statements in the source code, such as \texttt{logging.info()}, \texttt{printf()}~\cite{zhu2023loghublargecollectionlog} and other language-specific functions. Due to the diverse design goals of these systems, the information recorded in different log files is often difficult to cross-reference or consolidate, posing significant challenges for system administrators and engineers who need a unified view of system performance.

The growing volume of log data and the rapid expansion of log information make manual processing and analysis increasingly difficult. While logs offer valuable insights into the behavior and performance of complex computer systems, manual analysis is time-consuming, labor-intensive, and prone to errors. The need for quick troubleshooting, combined with the repetitive nature of log analysis, can lead to operator fatigue, burnout, and inconsistent interpretations, contributing to inefficiencies and inaccuracies in system diagnostics. Furthermore, logs often fail to provide real-time insights, making it harder to identify emerging issues or anomalies promptly.

We propose a unified framework, LogBabylon, to overcome these challenges. It adopts the semantic approach for log integration and leverages Large Language Models with the Retrieval Augmented Generation (RAG) ~\cite{lewis2021retrievalaugmentedgenerationknowledgeintensivenlp}. RAG enhances LLMs by enabling them to retrieve relevant, external information from databases or document repositories and generate contextually accurate responses. Combining internal and external resources, LLMs can significantly reduce the ``hallucination" and enhance the response accuracy.  

LLMs have demonstrated strong capabilities in different tasks like code translation ~\cite{10.1145/3664646.3664771},unit test generation~\cite{karanjai2024harnessing},  in understanding unstructured data~\cite{karanjai2024lookalike,karanjai2023smarter}, including logs, and can process vast amounts of information more efficiently than traditional semantic methods. However, LLMs alone are insufficient for real-time analysis, as they rely on pre-trained knowledge that may not account for the dynamic and evolving nature of security events. By integrating RAG, we create a system capable of retrieving relevant log data in real time and generating accurate, context-aware insights. These systems can be even extended with the help of confidential execution of these LLMs ~\cite{karanjai2024trusted}.

 LogBabylon leverages RAG and LLMs to overcome traditional semantic systems' scalability and contextual limitations. This enables more sophisticated and adaptable analysis of system activities and security events. LogBabylon significantly enhances situational awareness, improves the detection of complex attacks, and reduces the manual workload for system administrators and security analysts. As a result, it provides a more efficient and effective solution for modern security management, enhancing the capabilities of system administrators, engineers, and security analysts involved in log analysis and system diagnostics.

This research paper presents several key contributions to the field of log analysis, prompting a re-evaluation of current methodologies:

\begin{enumerate}[label=\textbf{(\arabic*)}]
    \item \textbf{Unified Framework:} How might LogBabylon's comprehensive approach to log analysis reshape our understanding of the classification, consolidation, and interpretation of diverse log formats within a single, cohesive system?
    
    \item \textbf{LLM-Powered Parsing:} In what ways does LogBabylon's utilization of Large Language Models (LLMs) for the accurate extraction of log templates minimize our reliance on manual intervention and domain-specific expertise?
    
    \item \textbf{RAG-Enhanced Analysis:} How does the integration of Retrieval Augmented Generation (RAG) technology allow LogBabylon to leverage a vast knowledge base of log examples, ultimately facilitating deeper insights and more accurate anomaly detection?
    
    \item \textbf{Variable-Aware Prompting and In-Context Learning:} Can incorporating advanced prompting techniques and in-context learning significantly enhance the LLM's understanding of log data and improve template extraction accuracy?
    
    \item \textbf{Human-Readable Insights:} How does LogBabylon's ability to generate clear, concise, and human-readable analyses empower users to understand their log data and make informed decisions?
\end{enumerate}

The structure of the paper will be as follows: Section 1 will be the introduction, setting the context for the study. Section 2 will cover related work, providing an overview of previous research in the field. Section 3 describes the methods and implementation used in the study, followed by Section 4, which focuses on the experiments conducted and result. Finally, Section 5 will conclude and summarize the key findings and the implications.

\section{Related Work}
Different approaches have been proposed to integrate log files. We examine them from three directions.

\subsection{Semantic Approach of Log Integration}
Traditional semantic approaches to log integration face significant challenges, primarily because they must handle log data from many different sources, each with its own format and structure ~\cite{cali2005comprehensive,ekelhart2018taming}. These methods often are not scalable enough to keep up with the massive amounts of log data generated by modern systems. Additionally, log events are highly specific to the context in which they occur, making it difficult to automate their interpretation. As a result, these limitations prevent semantic analysis from capturing key insights, especially when dealing with modern cyber threats and complex systems. 

SEPSES is a semantic log analysis framework that offers a platform for semantic-based security monitoring, auditing, and forensic investigations~\cite{EKELHART2018109}. It uses JSON-LD (JSON for Linking Data) to consolidate fragmented log information and extract and interlink related to security information. In this system, the log vocabulary stack is one of the key components that enables transforming the raw log data into a uniform RDF (Resource Description Framework) representation~\cite{pan2009resource}. The log vocabulary stack consists of two kinds of terms. One is the core vocabulary, \texttt{slog:core}, the foundation of the basic terms to describe log messages independent from their sources. The other one is the source-specific terms. To provide event specifications, they collect background knowledge, such as syslog events, Apache events, etc. 

The significant drawback of SPESES is that the log vocabulary and background knowledge are manually collected and added to the system. This limits the system's ability to interpret logs and link events dynamically. Thus, the scalability is a big concern for this system. As the volume of log event data surges, the semantic processing system must scale accordingly, which can be demanding for a system with a static design. Cyber threats constantly evolve, and semantic processing systems need continual updates to understand new patterns and threats. Keeping the system up-to-date requires ongoing effort and adaptation, which can be resource-intensive.

\subsection{AI Approach of Log Integration}
AI and machine learning have become increasingly central to integrating log data across different systems, offering enormous improvements in monitoring, security, and operational management efficiency and effectiveness. AI can automatically parse logs from different sources and normalize the data into a consistent format~\cite{8029742}. Machine learning models can learn from historical log data to understand normal system behavior. These model can detect anomalies or deviations in new log data, which could indicate potential issues like security breaches or system failures~\cite{chen2022experiencereportdeeplearningbased}. AI can analyze trends within log data to predict future system behaviors and potential problems. This proactive approach enables preemptive maintenance and security measures, minimizing downtime and preventing breaches before they occur~\cite{10386543}. 

There are numerous commercial and opensource tools for automatically analyzing log files, such as Semtext Logs~\cite{GIAMATTEI2024111906}, SolarWinds Loggly~\cite{SolarWindsLoggly}, Splunk~\cite{Splunk}, Rapid7 InsightOps~\cite{InsightOps}, and Sumo Logic~\cite{SumoLogic}. These tools offer centralized log management, allowing users to collect, store, and analyze logs from various sources, with most providing cloud-based solutions. There are approaches which look at LLM based solutions as well~\cite{gao2023retrieval}. They include search functionality and analytics features to help users gain insights from the data, as well as alerting systems to notify users of important events or anomalies. Additionally, they offer data visualization through charts, dashboards, or other graphical representations.

However, these tools often have drawbacks, such as being overly complex, offering limited log management capabilities, and performing poorly when handling large volumes of data or long-term retention. These difficulties highlight the trade-offs in log management tools, which are influenced by specific needs, technical expertise, and budget constraints. 

\subsection{Log Analysis}
The traditional approach to log analysis starts with log parsing, an essential step for preparing log data for further analysis. A notable method in this area is ``Drain'', an online log parsing technique designed to speed up this process~\cite{8029742}.
After parsing, the next step is log-based anomaly detection, which includes extracting relevant features and then using these features in a machine-learning model to identify and predict anomalous events~\cite{chen2022experiencereportdeeplearningbased}. 
These steps represent a foundational framework in the current landscape of log analysis technology.

However, traditional methods primarily focus on detecting anomalies. ``LogGPT''~\cite{10386543} is a novel approach in which a model is trained to predict the next log entry based on previous sequences. If an observed log key does not appear in the top 50\% of the model's prediction list, it is labeled as an anomaly. While this method utilizes large language models (LLMs) for anomaly detection, it does not address the system insight analysis.

By contrast, Xu et al. combined log parsing with text mining to analyze console logs and focused their research on detecting program bugs and runtime anomalies~\cite{10386543}. This approach, while innovative, remains limited to specific types of log analysis and does not encompass broader applications such as summarization or comprehensive system diagnostics.
The log analysis landscape is rapidly evolving, with AI and ML playing a central role in transforming how organizations handle and derive insights from their log data. The focus is on automation, real-time analysis, and making log data more accessible and actionable across different organizational roles. This actually gets emphasized by \cite{karlsen2024benchmarking} which shows the limitations on some of the approaches.

\section{Detailed Design of LogBabylon}
To overcome the complexities of log parsing, we introduce LogBabylon, an innovative approach powered by Large Language Models (LLMs). LogBabylon excels in accurately extracting log templates due to the LLMs' advanced comprehension abilities.  It strategically utilizes LLMs to optimize performance and minimize resource consumption. Designed for versatility, LogBabylon adapts to various log formats and domains while minimizing human intervention in tasks, such as data labeling and parameter tuning. Furthermore, it incorporates a dynamic feedback loop to refine the parsing granularity based on human input, ensuring accurate and efficient log analysis.

\begin{figure}[h!]
    \centering
    \includegraphics[width=1.0\linewidth]{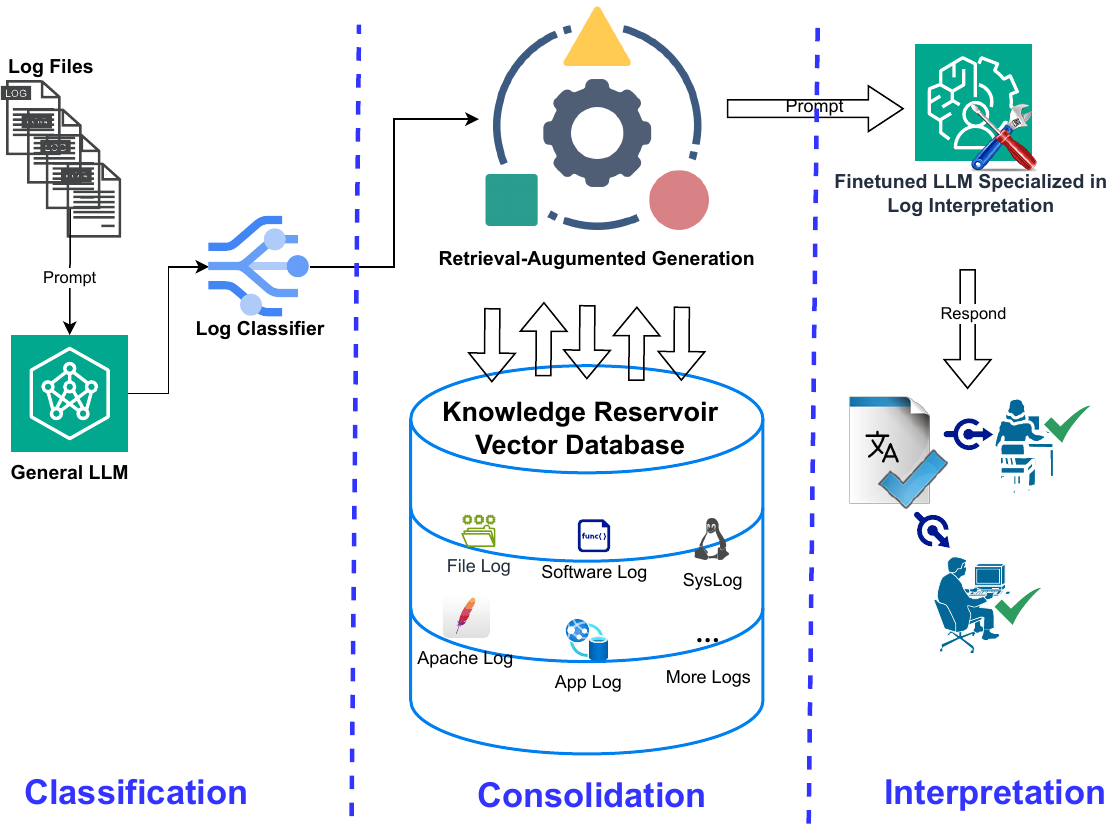}
    \caption{LogBabylon Architecture}
    \label{fig:architecture}
\end{figure}

The LogBabylon framework consists of a sequence of progressive steps: classification, consolidation, and interpretation. 
\figurename~\ref{fig:architecture} demonstrates high-level architecture design of LogBabylon.  

\subsection{Classification}
\subsubsection{Overview}
LogBabylon employs a streamlined pre-processing approach that minimizes the need for extensive domain expertise. Unlike methods that rely heavily on regular expressions to replace specific variables, such as IP addresses~\cite{he2017drain}, LogBabylon retains the original log message, allowing the LLM to fully grasp its context.  This approach, which uses plain space-based tokenization and leverages the LLM's native tokenizer~\cite{fu2022investigating,yu2023brain}, simplifies preprocessing while maintaining high log parsing efficiency.
\subsubsection{Algorithm}
LogBabylon's core algorithm centers on a prefix parse tree~\cite{he2017drain}. This tree structure efficiently matches incoming logs with existing clusters, thereby facilitating the rapid identification of similar entries. The algorithm intelligently determines when to invoke the LLM for template extraction, optimizing resource usage.  As the LLM identifies new templates, the tree dynamically updates to incorporate these findings, ensuring an accurate and evolving representation of the log data. This sophisticated approach enables LogBabylon to effectively process and analyze logs by combining the strengths of the prefix parse tree and the LLM.

\begin{figure}[h!]
    \centering
    \includegraphics[scale=0.5]{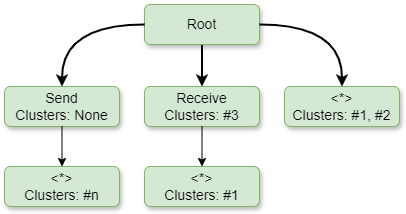}
    \caption{An example of the data structures in our method}
    \label{fig:logcluster}
\end{figure}

Three primary data structures form the backbone of our approach: a set of log clusters, a template pool, and a prefix parse tree. \figurename~\ref{fig:logcluster} visually represents this organizational structure. The subsequent discussion delineates their respective functionalities.

\textbf{Log Cluster:} A log cluster is a collection of logs with the same template. It tracks the individual log IDs and stores a log embedding created by an LLM encoder for future use. Each cluster is characterized by its log template, extracted via LLM, and possibly multiple syntax templates aiding the prefix tree in its traversal and template matching processes. While syntax templates correspond directly to the tokens of the raw logs, identifying static and variable parts, the log templates from the LLM may represent several tokens with a single placeholder. These syntax templates are stored in a dictionary, utilizing the token counts as keys and the corresponding template lists as values.

\textbf{Template Pool:} The template pool establishes a linkage, mapping log templates to their respective log clusters.

\textbf{Prefix Parse Tree:} In this tree structure, each node, except the root, represents a token. The wildcard token "<*>" is a universal matcher for any token. Importantly, the leaf nodes and nearly all nodes (except the root) can have pointers to log clusters that match the token sequence, starting from the root. A notable feature is that a single log cluster may be accessible from multiple nodes due to the possibility of having different syntax template variations for the same log cluster.

When a new log arrives, LogBabylon first breaks it into individual tokens. These tokens are then sequentially matched against the nodes in the prefix parse tree, gradually narrowing down the potential log clusters. The process continues until all tokens are matched or no more matches are found.

Unlike traditional methods that rely on similarity metrics and thresholds, LogBabylon employs a more definitive matching process, categorizing matches as strict, loose, or non-existent.  First, it checks whether the number of tokens in the incoming log aligns with the syntax templates of the cluster.  If the counts differ, a match is ruled out immediately.  Next, a "loose match" is attempted, where tokens from the syntax template and the log are aligned, allowing wildcard tokens ("<*>" ) to match any log token.  If a loose match is found, regular expressions are used to ensure strict alignment with the non-wildcard elements of the syntax template. A complete token alignment signifies a strict match.

The log is readily added to the corresponding cluster if a strict match is identified. However, if no strict match is found, LogBabylon calls upon its LLM-powered template extractor to generate a new template. The data structures are then updated accordingly, ensuring that the system continuously learns and adapts to the new log patterns. This dynamic approach allows LogBabylon to categorize logs accurately even when encountering previously unseen formats.

LogBabylon's accuracy in log parsing stems from the inherent capabilities of LLMs, eliminating the tedious task of fine-tuning the hyperparameters for different log sources.  LogBabylon simplifies the tuning process and minimizes the need for frequent LLM calls by leveraging the ability of LLMs to generate semantically accurate templates.  Theoretically, if the LLM consistently produces templates that perfectly reflect the actual structure of the logs, the number of LLM calls would be limited to the total number of distinct syntax templates. This roughly equals the total number of log templates, making LogBabylon a highly scalable solution.

The LogBabylon algorithm shown in Algorithm~\ref{alg:our} efficiently parses logs by combining template matching with the power of LLMs. It starts with a hierarchical tree structure and two key data structures: log\_clusters to store groups of similar logs, and template\_pool to store templates for quick retrieval.

\begin{algorithm}[tb]
   \caption{\textsc{LogBabylon}}
   \label{alg:our}
\begin{algorithmic}[1]
    \item[] {\bfseries Input:} $logs$, $root\_node \quad$  {\bfseries Result:} $log\_clusters$, $tree$
    \STATE $log\_clusters \leftarrow \{\}$
    \STATE $template\_pool \leftarrow \{\}$
    \STATE $partial\_match\_cache \leftarrow \{\}$  \quad \textit{// Cache for loose matches}
    \FOR{$log \textbf{ in } logs$}
        \STATE $matched\_clusters \gets \text{search}(tree, log)$
        \IF{strict\_match}
            \STATE $strict\_matched\_cluster.\text{add}(log)$
            \STATE $added \gets \text{True}$
        \ELSIF{loose\_match \textbf{or} no\_match}
            \STATE $template \gets \text{get\_llm\_template}(log, log\_clusters)$
            \IF{$template$ \textbf{in} $template\_pool$}
                \STATE $\text{update\_tree}(tree, log, template\_pool[template])$
                \STATE $added \gets \text{True}$
            \ELSE
                \STATE $loose\_matched\_clusters \gets 
                \textit{partial\_match\_cache}.\text{get}(log, [])$
                \IF{loose\_matched\_clusters = \{\}}
                    \STATE $loose\_matched\_clusters \gets \text{find\_loose\_matches}(log, log\_clusters)$
                    \STATE $partial\_match\_cache[log] \gets loose\_matched\_clusters$
                \ENDIF
                \FOR{$cluster$ \textbf{in} $loose\_matched\_clusters$}
                    \STATE check\_merge($log, cluster$)
                    \IF{merge}
                        \STATE $cluster.\text{update}(merged\_template)$
                        \STATE $cluster.\text{add}(log)$
                        \STATE $template\_pool[merged\_template] = cluster$
                        \STATE \textbf{break}
                        \STATE $added \gets \text{True}$
                    \ENDIF
                \ENDFOR
            \ENDIF
        \ENDIF
        \IF{\textbf{not} $added$}
            \STATE $new\_cluster \gets \text{create\_cluster}(log, template)$
            \STATE $\text{update\_tree}(tree, log, new\_cluster)$
            \STATE $template\_pool[template] = new\_cluster$
        \ENDIF
    \ENDFOR
\end{algorithmic}
\end{algorithm}

For each new log, LogBabylon searches for the tree for matching clusters. The log is immediately added to the cluster if a perfect (``strict'') match is found. Otherwise, the LLM generates a template for the log.

LogBabylon then checks if this template already exists in template\_pool. If so, the log is added to the corresponding cluster, and the tree structure is updated. If the template is new, then the algorithm attempts to merge the log with the most similar cluster. If the merge is successful, the cluster is updated with the new template and the log is added.  The template\_pool is also updated.

If no match or successful merge is found, a new cluster is created for the log and the tree is updated to include this new cluster. This process ensures the efficient grouping of logs based on both exact and approximate matches, leveraging the LLM's ability to extract meaningful templates and enhance clustering accuracy.

\subsection{Consolidation}
LogBabylon utilizes Retrieval Augmented Generation (RAG)~\cite{lewis2020retrieval} to enhance its log analysis capabilities. This approach ensures that it does not solely rely on internal knowledge but also leverages a vast collection of "normal" log entries to interpret new logs effectively.

Consider this collection of a comprehensive library of log examples that is efficiently stored in a vector database. When a new log entry arrives, LogBabylon attempts to match it with existing templates and queries this database to find similar past entries. 

The LLM then conducts a detailed semantic analysis by comparing the new log entry with the retrieved examples. This process allows LogBabylon to:
\begin{itemize}
    \item \textbf{Identify subtle patterns and anomalies:} By comparing the new log with known examples, LogBabylon can detect even minor deviations that may indicate unusual activity.
    \item \textbf{Understand the log context: } The retrieved examples offer valuable context, helping LogBabylon grasp the significance of the new log entry.
    \item \textbf{Generate more accurate interpretations:} The combination of internal knowledge with external examples enables LogBabylon to provide more accurate and insightful interpretations of log data.
\end{itemize}

This RAG-based approach renders LogBabylon a versatile and user-friendly tool for log analysis. Unlike traditional methods that require complex, multi-stage processing pipelines, LogBabylon offers an end-to-end solution that adapts seamlessly to any log source with a minimal configuration. By combining the power of LLMs with a rich knowledge base of log examples, LogBabylon simplifies log analysis and unlocks deeper insights from the data.

LogBabylon’s application of the Retrieval Augmented Generation (RAG) model can be divided into three steps:

\subsubsection{Finding Relevant Information}
When LogBabylon receives a new log entry, denoted by \(x\), it first seeks contextually relevant log entries from its knowledge base (stored in a vector database) to better understand the new entry. This step leverages dense-vector retrieval.

\begin{itemize}
    \item \textbf{Turning Logs into Vectors:} LogBabylon employs a pre-trained embedding model to convert each log entry into a dense vector representation~\cite{lee2019latent}. These vectors encapsulate the semantic meaning of the logs, enabling comparisons based on meaning rather than on literal words.
    
    \item \textbf{Calculating Similarity:} To identify relevant logs, LogBabylon compares the vector of the new log entry, \(x\), with the stored vectors in the database. This comparison uses the inner product, a mathematical operation that measures similarity between two vectors.
    
    \item \textbf{Retrieving the Best Matches:} The database returns the log entries whose vectors have the highest similarity scores to \(x\). This set of relevant log entries is denoted as \(z\).
\end{itemize}

\subsubsection{Generating a Response}
With the relevant context \(z\) retrieved for the new log entry \(x\), LogBabylon proceeds to generate a meaningful response, leveraging the Large Language Model (LLM), represented as \(f\). 

\begin{itemize}
    \item \textbf{Combining Input and Context:} The LLM takes both the new log entry \(x\) and the retrieved relevant logs \(z\) as input.
    
    \item \textbf{Generating Output:} Based on this combined input, the LLM generates a response, \(y\). This response can be an interpretation of the log, an anomaly detection flag, or another task-specific output.
\end{itemize}

The entire process can be expressed using the following equation:

\[
y = f(x, z)
\]

This equation illustrates how LogBabylon combines dense vector retrieval and LLMs to generate insightful outputs from log data. By leveraging both its knowledge base and the reasoning capabilities of the LLM, LogBabylon offers a simple yet powerful solution for log analysis.

\subsubsection{LLM Semantic Analysis in LogBabylon}

\begin{figure}[h!]
    \centering
    \includegraphics[scale=0.4]{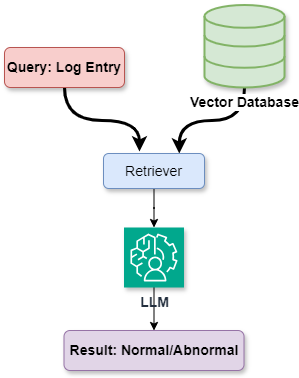}
    \caption{LLM Semantic Analysis in LogBabylon}
    \label{fig:rag}
\end{figure}

Once LogBabylon retrieves relevant log entries from its vector database, it must analyze and compare them with the new log entry. This is where the LLM’s semantic analysis capabilities come into play. \figurename~\ref{fig:rag} demonstrates this process.

\subsubsection*{Decoding the Vectors}
The retrieved vectors are first transformed back into their original log entry format. This is achieved using the decoder component of the embedding model, which reverses the encoding process, converting the abstract vector representation into human-readable log text.

\subsubsection*{Framing the Anomaly Detection Task}
LogBabylon frames the anomaly detection task as a question-answering problem. This approach leverages the natural ability of LLMs to interpret questions and provide human-like responses.

\subsubsection*{Crafting the Prompt}
A specialized prompt template is used to provide structured input to the LLM. The template includes:

\begin{itemize}
    \item \textbf{The new log entry:} The log that LogBabylon is analyzing.
    \item \textbf{The retrieved normal log entries:} These serve as context, offering a baseline for comparison.
    \item \textbf{The question:} The LLM is asked, \textit{“Is the new log entry normal or abnormal, given the provided examples of normal logs?”}
\end{itemize}

\subsubsection*{LLM Analysis}
The LLM processes this structured prompt by leveraging its language understanding and contextual reasoning abilities. It determines whether the new log entry deviates from the norm, producing one of the following outputs:

\begin{itemize}
    \item A simple classification: \textit{“normal”} or \textit{“abnormal.”}
    \item A detailed explanation: Describing the detected anomaly and its significance.
\end{itemize}

By framing anomaly detection as a question-answering task and harnessing the semantic capabilities of the LLM, LogBabylon can effectively identify unusual log entries, even those with subtle deviations from the expected behavior. This approach ensures accurate detection of potential issues and provides valuable insights into the log data.

\subsection{Interpretation}
LogBabylon's final step involves refining the insights gained from the previous stages and presenting them in a clear, human-readable format. This is where a specialized LLM, fine-tuned specifically for log interpretation, takes the center stage.

\subsubsection{Enhancing LLM Template Extraction}

While LogBabylon's base algorithm effectively matches log clusters, there is always room for improvement, particularly in the accuracy of the LLM template extractor. LogBabylon incorporates two key strategies to enhance this crucial component: variable aware prompting and in-context learning.

\subsubsection{Variable-Aware Prompting}
Inspired by research~\cite{li2023did} that emphasizes the importance of identifying and classifying variables within logs, LogBabylon employs variable-aware prompting. This technique encourages the LLM to not only identify which parts of the log message are variables, but also to categorize them into specific types\cite{li2023did}, such as timestamps, IP addresses, or error codes. This approach is akin to providing the LLM with a ``chain of thought''~\cite{wei2022chain} guiding it towards a deeper understanding of the log structure and meaning of its various components.

\subsubsection{In-Context Learning with K-Shot Demonstrations}
In-context learning (ICL) is a powerful technique that allows LLMs to learn new tasks without extensive fine-tuning~\cite{dong2022survey}. LogBabylon leverages ICL by providing the LLM with a few examples of log entries and their corresponding templates. These examples serve as ``demonstrations'' that guide the LLM towards generating accurate templates for new log entries.

\subsubsection{Combining the Power of Both}
LogBabylon seamlessly integrates variable-aware prompting and ICL to enhance template extraction. Each time the LLM is called upon to extract a template, it receives a prompt containing:

\begin{itemize}
    \item \textbf{Instructions:} A clear description of the template extraction task.
    \item \textbf{Demonstrations:} A small set of examples (\(k = 3\)) of log entries and their corresponding templates. These examples are carefully selected from a pool of previously extracted templates based on their similarity to the new log entry.
    \item \textbf{Seed Examples:} Ten examples representing different types of log parameters are included as initial seeds to guide the LLM in variable classification.
    \item \textbf{Query:} The new log entry for which the LLM needs to generate a template.
\end{itemize}

By combining clear instructions, relevant demonstrations, and seed examples, LogBabylon guides LLM to generate accurate and informative templates. This approach not only improves the accuracy of template extraction but also enhances the LLM's overall understanding of log data.

\subsubsection{Beyond Variable-Aware Prompting and ICL}
While variable-aware prompting and ICL significantly enhance LogBabylon's template extraction capabilities, there are other potential avenues for further improvement, including:

\begin{itemize}
    \item \textbf{Utilizing more powerful LLMs:} As the field of language models advances, LogBabylon can readily incorporate newer, more capable LLMs to further boost its performance.
    \item \textbf{Supervised fine-tuning:} Training the LLM on a labeled dataset of log entries and their corresponding templates can lead to even greater accuracy in template extraction.
\end{itemize}

\subsubsection{Human-Readable Output for Various Purposes}

The final output produced by LogBabylon is designed to be easily understood by humans. It can take various forms, depending on the specific needs of the user, including:

\begin{itemize}
    \item \textbf{Concise summaries of log events:} This provides a quick overview of what happened in the system.
    \item \textbf{Detailed explanations of anomalies:} This helps identify the root cause of problems and potential security threats.
    \item \textbf{Actionable insights for troubleshooting:} This guides users towards resolving issues and improving system performance.
\end{itemize}

This versatile output can be used for various purposes such as troubleshooting system errors, monitoring performance, detecting security breaches, and gaining a deeper understanding of user behavior. By providing precise, human-readable analyses, LogBabylon empowers users to make informed decisions and take appropriate actions based on their log data.

\subsection{Metrics for Evaluation}
We use existing metrics~\cite{zhong2024logparser} for our evaluation.

\textbf{Grouping Accuracy (GA) or Clustering Accuracy (CA):} This metric measures the ratio of log messages that are correctly grouped. 

\textbf{Parsing Accuracy (PA):} This metric evaluates the ability of the technique to extract templates accurately, a crucial aspect for tasks like anomaly detection. 

\textbf{F1 score of Grouping Accuracy (FGA):} This is a template-level metric that evaluates the fraction of correctly grouped templates. It uses the true number of templates ($N_g$), parsed templates ($N_p$), and correctly parsed templates ($N_c$) to calculate Precision ($P_{GA} = \frac{N_c}{N_p}$) and Recall ($R_{GA} = \frac{N_c}{N_g}$) of Grouping Accuracy. The F1 score is the harmonic mean of these two values.

\textbf{F1 score over template accuracy(FTA)}: 
FTA is the harmonic mean of Recall of Template Accuracy (RTA) and Precision of Template Accuracy (PTA). Similar to FGA, FTA evaluates correct template identification at the template level. A template is correct if log messages with the same parsed template share the same ground-truth template and the parsed template matches the ground-truth template exactly. 

\textbf{Precision Template Accuracy (PTA) }is a template-level metrics to evaluate the quality of parsing. PTA measures the ratio of correctly identified templates to the total number of identified templates. 

\textbf{Recall Template Accuracy (RTA)} measures the ratio of correctly identified templates to the total number of ground-truth templates. 


\section{Experiments}
To rigorously evaluate LogBabylon's performance, we utilize two datasets, loghub-2k \cite{zhu2023loghub} and logPub ~\cite{jiang2024large}, employing standard metrics like parsing accuracy alongside a novel "granularity distance" metric to assess the precision of template extraction.  Our analysis of loghub-2k focuses on illustrating the contribution of each design component, such as variable-aware prompting and in-context learning. Meanwhile, our evaluation with logPub aims to demonstrate the effectiveness and efficiency of our approach when applied to large-scale, real-world datasets.

\subsection{Testbed}

\subsubsection{Datasets}
We evaluate LogBabylon's efficacy using two comprehensive datasets: loghub-2k~\cite{zhu2023loghub} and logPub~\cite{jiang2024large}. Loghub-2k has established itself as a prominent benchmark in log parsing research, featuring a diverse collection of logs from 16 different systems. This dataset spans various computing environments including distributed systems, supercomputers, operating systems, mobile platforms, server applications, and standalone software packages5. Each system in Loghub-2k is represented by 2,000 carefully annotated log messages, thereby providing a solid foundation for evaluating parsing techniques.

\subsubsection{Experimental Environment}
Experiments were conducted on an Ubuntu 20.04.3 LTS server equipped with 512GB of RAM. We employed ChatGPT (gpt-3.5-turbo-0301) and GPT-4 (gpt-4-0613) for template extraction, accessed via the official OpenAI API. Log embedding was performed using the text-embedding-ada-002 model~\cite{OpenAI-ada}.


The LLM models used in the experiment are configured in the following way:
\begin{itemize}
    \item \textbf{Template Extraction:} ChatGPT (gpt-3.5-turbo-0301), GPT-4 (gpt-4-0613)
    \item \textbf{Log Embedding:} text-embedding-ada-002
    \item \textbf{Fine-tuning:} Gemma2-9b~\cite{team2024gemma}
    \item \textbf{Temperature:} 0. The temperature parameter is set to 0 to minimize the variability of the LLM outputs.
\end{itemize}

\subsubsection{In-context Learning}
For in-context learning, 32 log-template pairs were uniformly sampled from the first 10\% of each dataset based on token length, serving as candidate logs and fine-tuning examples.

\begin{equation}
\begin{aligned}
\text{Sample Size} &= 32 \text{ log-template pairs} \\
\text{Sampling Method} &= \text{Uniform from first 10\% of dataset}
\end{aligned}
\end{equation}

This experimental setup ensures a consistent and reproducible environment for evaluating log parsing techniques using state-of-the-art language models and embedding approaches.

\subsection{Results}

\subsubsection{Loghub-2k}
The primary focus of our experiments with the loghub-2k dataset was twofold: first, to rigorously evaluate the individual contributions of LogBabylon's core components, and second, to utilize this dataset as a development set for refining various aspects of the system. This includes optimizing the prompts used for LLM-driven template extraction, fine-tuning the criteria for cluster merging, and enhancing the overall verification process.

\begin{table}[]
\caption{Comparison with existing methods in loghub}
    \label{tab:LLMtable}
\resizebox{\columnwidth}{!}{%
\begin{tabular}{|l|l|c|c|}
\hline
                                 & \# of labeled logs & Clustering Accuracy & Parsing Accuracy \\ \hline
\multirow{2}{*}{Eval of chatgpt} & 0                  & 72.1                & 54.3             \\ \cline{2-4} 
                                 & 4                  & 76.1                & 79.0             \\ \hline
DivLog\cite{xu2023prompting}     & 200                & 92.8                & 98.1             \\ \hline
\multirow{3}{*}{LogBabylon}      & 0                  & 88.9                & 79.1             \\ \cline{2-4} 
                                 & 4                  & 82.2                & 63.9             \\ \cline{2-4} 
                                 & 200                & 97.2                & 94.0             \\ \hline
\end{tabular}%
}

\end{table}

\subsubsection{Comparison with Existing Parsers based on LLM}
Evaluating existing LLM-based log parsers, which typically process logs line-by-line, poses significant challenges when applied to the extensive LogPub dataset due to the substantial computational costs associated with numerous LLM API calls. To address this issue, we conducted a comparative analysis using the more manageable Loghub-2k dataset. However, it is essential to acknowledge the limitations of this smaller dataset; its restricted scope means that a carefully selected set of labeled samples could potentially cover a large proportion of the log templates, which may lead to overly optimistic performance metrics that do not generalize well to larger, more diverse log collections. Despite these constraints, our method, LogBabylon, achieves competitive or superior performance compared with existing approaches when utilizing an equivalent number of labeled logs, as illustrated in Table~\ref{tab:LLMtable}. This outcome highlights LogBabylon's effective use of Large Language Models for log parsing tasks within the confines of a smaller dataset. It suggests that our approach offers tangible benefits in accuracy and efficiency. Furthermore, the strong performance indicates promise for future applications to larger and more complex log datasets, providing valuable insights into the efficacy of our LLM-based log parsing strategy while laying the groundwork for more extensive assessments down the line.

\subsubsection{LogPub}
The results from our evaluation of the extensive logPub dataset in Table~\ref{tab:logpub} demonstrate LogBabylon's strong performance. Even without any fine-tuning for specific log formats, LogBabylon significantly outperforms all other methods in key metrics like Grouping Accuracy (GA), Full Accuracy (FGA), and Parsing Accuracy (PA). 

While LogBabylon's overall accuracy (PA) is slightly lower, this is mainly due to the complexities of perfectly matching the varying levels of detail in different log formats.  LogBabylon shows consistent performance across all 14 datasets within logPub without needing any adjustments for each specific log type. This highlights its adaptability and generalizability.

When we incorporate in-context learning (ICL) to calibrate LogBabylon for specific log formats (LogBabylon-C), we see a further improvement, particularly in template parsing metrics like PA and FTA. This demonstrates how ICL helps bridge the gap between general log parsing and the nuances of individual log structures.

\begin{table}[]
\caption{Comparison on the logPub dataset}
    \label{tab:logpub}
\resizebox{\columnwidth}{!}{%
\begin{tabular}{|c|cccccc|cccccc|cccccc|}
\hline
 &
  \multicolumn{6}{c|}{Drain} &
  \multicolumn{6}{c|}{Uniparser} &
  \multicolumn{6}{c|}{LogBabylon} \\ \hline
 &
  \multicolumn{1}{c|}{GA} &
  \multicolumn{1}{c|}{PA} &
  \multicolumn{1}{c|}{FGA} &
  \multicolumn{1}{c|}{FTA} &
  \multicolumn{1}{c|}{GGD} &
  PGD &
  \multicolumn{1}{c|}{GA} &
  \multicolumn{1}{c|}{PA} &
  \multicolumn{1}{c|}{FGA} &
  \multicolumn{1}{c|}{FTA} &
  \multicolumn{1}{c|}{GGD} &
  PGD &
  \multicolumn{1}{c|}{GA} &
  \multicolumn{1}{c|}{PA} &
  \multicolumn{1}{c|}{FGA} &
  \multicolumn{1}{c|}{FTA} &
  \multicolumn{1}{c|}{GGD} &
  PGD \\ \hline
Proxifier &
  \multicolumn{1}{c|}{69.2} &
  \multicolumn{1}{c|}{68.8} &
  \multicolumn{1}{c|}{20.6} &
  \multicolumn{1}{c|}{17.6} &
  \multicolumn{1}{c|}{4} &
  14 &
  \multicolumn{1}{c|}{50.9} &
  \multicolumn{1}{c|}{63.4} &
  \multicolumn{1}{c|}{28.6} &
  \multicolumn{1}{c|}{45.7} &
  \multicolumn{1}{c|}{5} &
  10 &
  \multicolumn{1}{c|}{44.9} &
  \multicolumn{1}{c|}{55.8} &
  \multicolumn{1}{c|}{35.2} &
  \multicolumn{1}{c|}{46.9} &
  \multicolumn{1}{c|}{4} &
  8 \\ \hline
Linux &
  \multicolumn{1}{c|}{\textbf{68.6}} &
  \multicolumn{1}{c|}{11.1} &
  \multicolumn{1}{c|}{77.8} &
  \multicolumn{1}{c|}{25.9} &
  \multicolumn{1}{c|}{30} &
  432 &
  \multicolumn{1}{c|}{28.5} &
  \multicolumn{1}{c|}{16.4} &
  \multicolumn{1}{c|}{45.1} &
  \multicolumn{1}{c|}{23.2} &
  \multicolumn{1}{c|}{108} &
  274 &
  \multicolumn{1}{c|}{23.8} &
  \multicolumn{1}{c|}{14.3} &
  \multicolumn{1}{c|}{70.5} &
  \multicolumn{1}{c|}{41.0} &
  \multicolumn{1}{c|}{16} &
  71 \\ \hline
Apache &
  \multicolumn{1}{c|}{\textbf{100.0}} &
  \multicolumn{1}{c|}{72.7} &
  \multicolumn{1}{c|}{100.0} &
  \multicolumn{1}{c|}{51.7} &
  \multicolumn{1}{c|}{\textbf{0}} &
  21 &
  \multicolumn{1}{c|}{94.8} &
  \multicolumn{1}{c|}{94.2} &
  \multicolumn{1}{c|}{68.7} &
  \multicolumn{1}{c|}{26.9} &
  \multicolumn{1}{c|}{11} &
  31 &
  \multicolumn{1}{c|}{88.0} &
  \multicolumn{1}{c|}{75.4} &
  \multicolumn{1}{c|}{88.0} &
  \multicolumn{1}{c|}{57.6} &
  \multicolumn{1}{c|}{0} &
  7 \\ \hline
Zookeeper &
  \multicolumn{1}{c|}{99.4} &
  \multicolumn{1}{c|}{84.3} &
  \multicolumn{1}{c|}{90.4} &
  \multicolumn{1}{c|}{61.4} &
  \multicolumn{1}{c|}{2} &
  30 &
  \multicolumn{1}{c|}{98.8} &
  \multicolumn{1}{c|}{\textbf{98.8}} &
  \multicolumn{1}{c|}{66.1} &
  \multicolumn{1}{c|}{51.0} &
  \multicolumn{1}{c|}{14} &
  31 &
  \multicolumn{1}{c|}{86.9} &
  \multicolumn{1}{c|}{72.1} &
  \multicolumn{1}{c|}{75.9} &
  \multicolumn{1}{c|}{63.7} &
  \multicolumn{1}{c|}{2} &
  17 \\ \hline
Hadoop &
  \multicolumn{1}{c|}{92.1} &
  \multicolumn{1}{c|}{54.1} &
  \multicolumn{1}{c|}{78.5} &
  \multicolumn{1}{c|}{38.4} &
  \multicolumn{1}{c|}{18} &
  210 &
  \multicolumn{1}{c|}{69.1} &
  \multicolumn{1}{c|}{88.9} &
  \multicolumn{1}{c|}{62.8} &
  \multicolumn{1}{c|}{47.6} &
  \multicolumn{1}{c|}{38} &
  119 &
  \multicolumn{1}{c|}{82.5} &
  \multicolumn{1}{c|}{59.5} &
  \multicolumn{1}{c|}{76.8} &
  \multicolumn{1}{c|}{48.4} &
  \multicolumn{1}{c|}{12} &
  95 \\ \hline
HealthApp &
  \multicolumn{1}{c|}{86.2} &
  \multicolumn{1}{c|}{31.2} &
  \multicolumn{1}{c|}{1.0} &
  \multicolumn{1}{c|}{0.4} &
  \multicolumn{1}{c|}{11} &
  138 &
  \multicolumn{1}{c|}{46.1} &
  \multicolumn{1}{c|}{81.7} &
  \multicolumn{1}{c|}{74.5} &
  \multicolumn{1}{c|}{46.2} &
  \multicolumn{1}{c|}{16} &
  60 &
  \multicolumn{1}{c|}{87.8} &
  \multicolumn{1}{c|}{51.2} &
  \multicolumn{1}{c|}{84.1} &
  \multicolumn{1}{c|}{72.0} &
  \multicolumn{1}{c|}{4} &
  4 \\ \hline
OpenStack &
  \multicolumn{1}{c|}{75.2} &
  \multicolumn{1}{c|}{2.9} &
  \multicolumn{1}{c|}{0.7} &
  \multicolumn{1}{c|}{0.2} &
  \multicolumn{1}{c|}{6618} &
  23** &
  \multicolumn{1}{c|}{\textbf{100.0}} &
  \multicolumn{1}{c|}{51.6} &
  \multicolumn{1}{c|}{96.9} &
  \multicolumn{1}{c|}{28.9} &
  \multicolumn{1}{c|}{1} &
  7 &
  \multicolumn{1}{c|}{88.0} &
  \multicolumn{1}{c|}{43.6} &
  \multicolumn{1}{c|}{88.0} &
  \multicolumn{1}{c|}{69.7} &
  \multicolumn{1}{c|}{0} &
  10 \\ \hline
HPC &
  \multicolumn{1}{c|}{79.3} &
  \multicolumn{1}{c|}{72.1} &
  \multicolumn{1}{c|}{30.9} &
  \multicolumn{1}{c|}{15.2} &
  \multicolumn{1}{c|}{10} &
  178 &
  \multicolumn{1}{c|}{77.7} &
  \multicolumn{1}{c|}{94.1} &
  \multicolumn{1}{c|}{66.0} &
  \multicolumn{1}{c|}{35.1} &
  \multicolumn{1}{c|}{10} &
  58 &
  \multicolumn{1}{c|}{76.0} &
  \multicolumn{1}{c|}{82.9} &
  \multicolumn{1}{c|}{66.9} &
  \multicolumn{1}{c|}{63.9} &
  \multicolumn{1}{c|}{5} &
  158 \\ \hline
Mac &
  \multicolumn{1}{c|}{76.1} &
  \multicolumn{1}{c|}{35.7} &
  \multicolumn{1}{c|}{22.9} &
  \multicolumn{1}{c|}{6.9} &
  \multicolumn{1}{c|}{102} &
  1347 &
  \multicolumn{1}{c|}{73.7} &
  \multicolumn{1}{c|}{68.8} &
  \multicolumn{1}{c|}{69.9} &
  \multicolumn{1}{c|}{28.3} &
  \multicolumn{1}{c|}{73} &
  624 &
  \multicolumn{1}{c|}{78.9} &
  \multicolumn{1}{c|}{26.7} &
  \multicolumn{1}{c|}{74.5} &
  \multicolumn{1}{c|}{31.9} &
  \multicolumn{1}{c|}{37} &
  391 \\ \hline
OpenSSH &
  \multicolumn{1}{c|}{70.7} &
  \multicolumn{1}{c|}{58.6} &
  \multicolumn{1}{c|}{87.2} &
  \multicolumn{1}{c|}{48.7} &
  \multicolumn{1}{c|}{3} &
  33 &
  \multicolumn{1}{c|}{27.5} &
  \multicolumn{1}{c|}{28.9} &
  \multicolumn{1}{c|}{0.9} &
  \multicolumn{1}{c|}{0.5} &
  \multicolumn{1}{c|}{15} &
  26 &
  \multicolumn{1}{c|}{68.6} &
  \multicolumn{1}{c|}{60.7} &
  \multicolumn{1}{c|}{84.6} &
  \multicolumn{1}{c|}{77.7} &
  \multicolumn{1}{c|}{1} &
  8 \\ \hline
Spark &
  \multicolumn{1}{c|}{88.8} &
  \multicolumn{1}{c|}{39.4} &
  \multicolumn{1}{c|}{86.1} &
  \multicolumn{1}{c|}{41.2} &
  \multicolumn{1}{c|}{18} &
  239 &
  \multicolumn{1}{c|}{85.4} &
  \multicolumn{1}{c|}{79.5} &
  \multicolumn{1}{c|}{2.0} &
  \multicolumn{1}{c|}{1.2} &
  \multicolumn{1}{c|}{62} &
  186 &
  \multicolumn{1}{c|}{85.9} &
  \multicolumn{1}{c|}{70.6} &
  \multicolumn{1}{c|}{75.0} &
  \multicolumn{1}{c|}{40.7} &
  \multicolumn{1}{c|}{14} &
  130 \\ \hline
Thunderbird &
  \multicolumn{1}{c|}{83.1} &
  \multicolumn{1}{c|}{21.6} &
  \multicolumn{1}{c|}{23.7} &
  \multicolumn{1}{c|}{7.1} &
  \multicolumn{1}{c|}{137} &
  2043 &
  \multicolumn{1}{c|}{57.9} &
  \multicolumn{1}{c|}{65.4} &
  \multicolumn{1}{c|}{68.2} &
  \multicolumn{1}{c|}{29.0} &
  \multicolumn{1}{c|}{194} &
  976 &
  \multicolumn{1}{c|}{64.2} &
  \multicolumn{1}{c|}{50.2} &
  \multicolumn{1}{c|}{70.4} &
  \multicolumn{1}{c|}{49.3} &
  \multicolumn{1}{c|}{92} &
  583 \\ \hline
BGL &
  \multicolumn{1}{c|}{91.9} &
  \multicolumn{1}{c|}{40.7} &
  \multicolumn{1}{c|}{62.4} &
  \multicolumn{1}{c|}{19.3} &
  \multicolumn{1}{c|}{48} &
  434 &
  \multicolumn{1}{c|}{91.8} &
  \multicolumn{1}{c|}{94.9} &
  \multicolumn{1}{c|}{62.4} &
  \multicolumn{1}{c|}{21.9} &
  \multicolumn{1}{c|}{43} &
  209 &
  \multicolumn{1}{c|}{82.5} &
  \multicolumn{1}{c|}{71.3} &
  \multicolumn{1}{c|}{69.4} &
  \multicolumn{1}{c|}{44.0} &
  \multicolumn{1}{c|}{30} &
  136 \\ \hline
HDFS &
  \multicolumn{1}{c|}{99.9} &
  \multicolumn{1}{c|}{62.1} &
  \multicolumn{1}{c|}{93.5} &
  \multicolumn{1}{c|}{60.9} &
  \multicolumn{1}{c|}{2} &
  6 &
  \multicolumn{1}{c|}{\textbf{100.0}} &
  \multicolumn{1}{c|}{94.8} &
  \multicolumn{1}{c|}{96.8} &
  \multicolumn{1}{c|}{58.1} &
  \multicolumn{1}{c|}{1} &
  1 &
  \multicolumn{1}{c|}{88.0} &
  \multicolumn{1}{c|}{83.4} &
  \multicolumn{1}{c|}{65.7} &
  \multicolumn{1}{c|}{50.9} &
  \multicolumn{1}{c|}{4} &
  23 \\ \hline
\end{tabular}%
}

\end{table}

\subsubsection{Different LLMs}
LogBabylon is designed to work with a variety of language models, allowing for flexibility and adaptability. This evaluation specifically aimed to explore how different LLMs impact its performance and efficiency.  Our findings are summarized in \tablename~\ref{tab:diffllm}.

\begin{table}[]
\caption{LogBabylon with different LLM}
    \label{tab:diffllm}
\resizebox{\columnwidth}{!}{%
\begin{tabular}{|l|c|ccc|ccccc|c|}
\hline
 &
  Avg. \# of &
  \multicolumn{3}{c|}{Avg. Time(s)} &
  \multicolumn{5}{c|}{Avg. Metrics} &
   \\ \hline
 &
  LLM Calls &
  \multicolumn{1}{c|}{Per Infer.} &
  \multicolumn{1}{c|}{Base} &
  Total &
  \multicolumn{1}{c|}{GA} &
  \multicolumn{1}{c|}{PA} &
  \multicolumn{1}{c|}{FGA} &
  \multicolumn{1}{c|}{FTA} &
  GGD &
  PGD \\ \hline
LogBabylon w/ GPT-3.5-turbo &
  621.5 &
  \multicolumn{1}{c|}{0.57} &
  \multicolumn{1}{c|}{564.7} &
  893.3 &
  \multicolumn{1}{c|}{88.6} &
  \multicolumn{1}{c|}{69.7} &
  \multicolumn{1}{c|}{87.3} &
  \multicolumn{1}{c|}{62.4} &
  29.4 &
  220.8 \\ \hline
LogBabylon with GPT-4 &
  468.8 &
  \multicolumn{1}{c|}{4.62} &
  \multicolumn{1}{c|}{498.3} &
  2461.6 &
  \multicolumn{1}{c|}{98.2} &
  \multicolumn{1}{c|}{74.5} &
  \multicolumn{1}{c|}{93.4} &
  \multicolumn{1}{c|}{69.4} &
  19.3 &
  146.4 \\ \hline
LogBabylon w/ fine-tuned Gemma2-9b(32shot) &
  7282.3 &
  \multicolumn{1}{c|}{2.79} &
  \multicolumn{1}{c|}{683.5} &
  2499.9 &
  \multicolumn{1}{c|}{85.6} &
  \multicolumn{1}{c|}{78.5} &
  \multicolumn{1}{c|}{72.8} &
  \multicolumn{1}{c|}{54.4} &
  50.5 &
  213.4 \\ \hline
\end{tabular}%
}

\end{table}

\subsection{Threats to Validity}
Ensuring the reliability of our evaluation was a key priority. We tackled potential issues like the LLM simply memorizing templates from its training data by showing the clear improvement brought about by in-context learning.  To keep things consistent and fair, we used the same LLM (gpt-turbo-3.5-0613) as in related research.  We also minimized unpredictable variations in the LLM's output by fixing the "temperature" setting and running each experiment multiple times, averaging the results.  Finally, to avoid any bias from how we set up the experiments, we compared LogBabylon to other leading methods using their best settings and publicly available code, all within the same testing environment. This ensured our results were in line with what others have reported.

\section{Conclusion}
In conclusion, this research paper introduces LogBabylon, a unified framework that effectively addresses the challenges of integrating and analyzing diverse log data. By harnessing the capabilities of Large Language Models (LLMs) and Retrieval Augmented Generation (RAG) technology, LogBabylon offers a robust and versatile solution for log analysis. Our key contributions include a unified framework that provides a comprehensive approach to log analysis, encompassing classification, consolidation, and interpretation of diverse log formats within a single, cohesive system. Additionally, LogBabylon leverages LLMs to accurately extract log templates, minimizing the need for manual intervention and domain-specific expertise. The integration of RAG technology enables LogBabylon to utilize a vast knowledge base of log examples, facilitating deeper insights and more accurate anomaly detection. Furthermore, it incorporates advanced variable-aware prompting techniques and in-context learning to enhance the LLM's understanding of log data and improve template extraction accuracy. LogBabylon generates clear, concise, and human-readable analyses, empowering users to understand their log data and make informed decisions. Through rigorous evaluation on the loghub-2k and logPub datasets, we demonstrate LogBabylon's superior performance compared to existing methods. Its adaptability, accuracy, and efficiency make it a valuable tool for various applications, including troubleshooting, performance monitoring, and anomaly detection. Future research directions include exploring the integration of more powerful LLMs, expanding the knowledge base with diverse log examples, and developing interactive interfaces for user feedback and granularity calibration. By continuously refining and enhancing LogBabylon, we aim to provide an even more robust and user-friendly solution for comprehensive log analysis.



\section*{Acknowledgments}
  This work was supported by the US National Security Agency (H98230-22-1-0323) and part of the pipeline creation was supported by cloud credits from Google Developer Expert program. 
  
  The authors would like to thank the anonymous reviewers for their helpful comments and feedback, which greatly strengthened the final version of the paper.
  
\noindent\insertcreditsstatement

\balance
\bibliographystyle{ACM-Reference-Format}
\bibliography{logbabylon-bibliography} 

\end{document}